\documentclass[useAMS,usenatbib]{mn2e}

\include{epsf}
\newcommand{\be}{\begin{equation}}
\newcommand{\ee}{\end{equation}}
\newcommand{\ba}{\begin{eqnarray}}
\newcommand{\ea}{\end{eqnarray}}
\newcommand{\nn}{\nonumber}
\newcommand{\logten}{{\rm log}_{10}}
\def\simless{\mathbin{\lower 3pt\hbox
   {$\rlap{\raise 4pt\hbox{$\char'074$}}\mathchar"7218$}}}
\def\simgreat{\mathbin{\lower 3pt\hbox
   {$\rlap{\raise 4pt\hbox{$\char'076$}}\mathchar"7218$}}}   

\title[Measuring luminosity function evolution]
{A new method to measure evolution of the galaxy luminosity function} 
\author[Dye \& Eales]  
{S. Dye$^1$\thanks{E-mail: s.dye@astro.cf.ac.uk} \& S. A. Eales$^1$
\vspace{2mm}\\
$^1$School of Physics \& Astronomy, Cardiff University, Queens Buildings,
The Parade, Cardiff, CF24 3AA, U.K.}

\begin{document}

\date{Accepted by MNRAS.}

\pagerange{\pageref{firstpage}--\pageref{lastpage}} \pubyear{2009}

\maketitle

\label{firstpage}

\begin{abstract}
We present a new efficient technique for measuring evolution of the
galaxy luminosity function. The method reconstructs the evolution over
the luminosity-redshift plane using any combination of three input
dataset types: 1) number counts, 2) galaxy redshifts, 3) integrated
background flux measurements. The evolution is reconstructed in
adaptively sized regions of the plane according to the input data as
determined by a Bayesian formalism. We demonstrate the performance of
the method using a range of different synthetic input datasets. We
also make predictions of the accuracy with which forthcoming surveys
conducted with SCUBA2 and the Herschel Space Satellite will be able to
measure evolution of the sub-millimetre luminosity function using the
method.
\end{abstract}

\begin{keywords}
Galaxies: evolution -- Methods: statistical
\end{keywords}

\section{Introduction}

A cornerstone of any working model of the formation of structure in
the universe is knowledge of the galaxy luminosity function (GLF).
The GLF is a measure of the comoving space density of galaxies per
interval in luminosity. Determining how the GLF changes with
cosmological epoch provides far-reaching insights into the
processes that dictate the means by which galaxies form and
evolve. This places strong statistical constraints on evolutionary
theories, enabling determination of key characteristics such as the
epoch of galaxy formation, merger rates, transformations between
population types and the history of the universe's global rate of
formation of stars.

The value of measuring the evolution of GLFs has been appreciated for
several decades and as such, many techniques have been employed to
achieve this aim. Without any loss of generality, the GLF can be
expressed as the local GLF scaled by an evolution function. The
simplest and most direct method of estimating this evolution function
is a model independent one where the GLF is measured at different
epochs and compared with the local GLF. This provides a set of
discrete estimates of the evolution at specific epochs. However,
adopting a model-based procedure yields significant advantages. A
well-proven approach is to assume the evolution function depends only
on luminosity and redshift (see next section). For a given model of
the evolution function chosen a priori, the best fit model is
determined, subject to a set of observational constraints which
contribute to different regions in the luminosity-redshift ($L-z$)
plane. The virtue of model-based techniques is that a model consistent
with all available observations can be used to make predictions by
extrapolation into regions where data are sparse or lacking.
Furthermore, a model can be used to identify those datasets whose
improvement would most efficiently increase the constraints on the
evolution function.

Model-based techniques are typically regarded as belonging to one of
two types. Parametric evolution models adhere to some preconceived
notion regarding the evolution of galaxies, for example, pure
luminosity evolution or luminosity + density evolution
\citep[see][and references therein for recent examples]{wall08}. 
Although an advantage of this type of modelling is that smooth
functions are readily extrapolated, a major disadvantage is that the
real evolution function may take on a very different form. So called
'free-form' methods are not biased in this way and allow for a 
greater degree of flexibility over the $L-z$ plane when attempting to
determine the evolution function.

Freeform techniques date back over three decades.
\citet{robertson78,robertson80} developed an iterative freeform method
to evaluate the evolution of the luminosity function for radio
galaxies. The method was limited by allowing the evolution to vary
only as a function of redshift (and not luminosity) and not giving a
satisfactory indication of the range of solutions permitted by the
observations. \citet{peacock81} introduced a new method that
incorporated luminosity dependence and measured the uncertainty on the
evolution by considering the variation between different freeform
model predictions. This improved technique saw successful application
\citep[e.g.,][]{peacock85,dunlop90} although by the authors' own
admission, the full extent of the uncertainty still could not be
verified by the range of models tested. Furthermore, the evolution
function was modelled as a series expansion and the set of best-fit
expansion coefficients determined by minimising $\chi^2$ using a
non-linear search. This is both inefficient and does not guarantee
that the global minimum has been found.

In this paper, we propose a new method, inspired by Peacock \& Gull
but offering some significant improvements. The method computes the
evolution as a discretised (i.e., pixellised) function over the $L-z$.
plane. This has the advantage that the evolution can be solved
linearly, ensuring the best fit solution is always found with a
single, highly efficient matrix inversion. Crucially, the full range
of solutions permitted by the observations can be determined in a few
simple extra computational steps. Despite being pixellised, we
demonstrate how linear regularisation allows the evidence to be
extrapolated into regions of the $L-z$ plane that are lacking data or
where data are sparse. Finally, we show how the pixellisation can be
adapted according to the constraints provided by the data.

The original motivation for this work was to investigate evolution of
the luminosity function in the sub-millimetre (submm) waveband.  The
far-infrared (far-IR) and submm wavebands are particularly important
for investigating the evolution of galaxies, not least because
approximately half of the energy emitted by stars and active galactic
nuclei since the big bang has been absorbed by dust and then
re-radiated in these wavebands \citep{fixsen98}. In view of this, our
interests lie in investigating not only the monochromatic luminosity
functions at these wavelengths, but also the `dust luminosity
function', i.e., the space density of galaxies as a function of the
total luminosity emitted by the dust in a galaxy.

Presently, observational data in the submm are particularly
poor. However, this situation will begin to rapidly improve with the
arrival of new submm instruments such as the second generation Submm
Common User Bolometer Array (SCUBA2) and the Herschel Space
Observatory ({\sl Herschel}). Although the method we have developed is
applicable to the luminosity function in any waveband, we describe and
explore its use in this paper with an eye on its future application to
forthcoming submm data.

In section \ref{sec_method} we present the method. Section
\ref{sec_demo} demonstrates the method with a range of synthetic
datasets providing differing levels of constraints on the $L-z$
plane. Finally, we summarise in section \ref{sec_summary} and discuss
practical aspects of applying the method to real data.

\section{The method}
\label{sec_method}

The overall aim of the method is to provide an estimate of the
evolution of the GLF. We therefore write the GLF, $\phi$, as the
product of the local GLF, $\phi_0$ and an evolution function, $E$:
\be
\label{eq_lf_evol}
\phi(L,z) = \phi_0(L)\,E(L,z) \, .
\ee
We have assumed that $E$ is a bivariate function of total
far-IR/submm luminosity, $L$, and redshift, $z$. We work in terms of
the bolometric luminosity rather than a monochromatic luminosity, in
keeping with the traditions adopted in submm astronomy.

Until very recently, there were no direct submm measurements of $\phi_0$
\citep[see][]{eales09}. Instead, the local GLF was estimated
by extrapolating from shorter wavelength observations. In this way,
the measurements had small statistical errors but, because of the
uncertainty in the extrapolation, potentially large systematic
errors. In what follows, we work towards determining $E$ directly,
accepting the fact that in application to real submm data, $E$
potentially has systematic errors because of the uncertainty in 
$\phi_0$. Of course, it is a trivial exercise to re-write the
formalism below in terms of $\phi(L,z)$ instead of $E(L,z)$ and use
$\phi_0$ as an observational constraint. However, in order to
determine $E$ from $\phi(L,z)$ obtained in this way, one would have to
renormalise by $\phi_0$. This would therefore introduce the exact
same systematic uncertainty on $E$ as that encountered by computing
$E$ directly.

The crux of the method is to write $E$ as a discrete function, instead
of the continuous form it has above. In this way, the evolution
function takes on discrete values, $e_j$, in pixels which span the
$L-z$ plane. As we show below, this discretisation turns the problem
of determining $E$ into a more desirable linear one.

Observational constraints on $E$ can be categorised into three
different types: number counts, galaxy redshifts and integrated
background flux. Number count data provide the number of detected
galaxies per unit area of sky binned by flux. The number of galaxies
between two flux limits gives information about the integral of the
luminosity function over an extended region of the $L-z$ plane, but
does not provide information about whether the sources are luminous
galaxies at high redshift or galaxies with low luminosity at low
redshift in that region. Writing this in terms of the luminosity
function, the expected number of detectable galaxies within a given
flux bin $i$ is
\be
\label{eq_ncounts}
n_i=\int^{z=\infty}_{z=0} {\rm d}V(z) 
\int^{L^i_u(z)}_{L^i_l(z)} {\rm d}L \, \phi(L,z) 
\ee
where the lower and upper luminosity limits $L^i_l$ and $L^i_u$
correspond to the lower and upper flux limits of the bin and are
therefore a function of redshift. These limits also depend on the
frequency at which the flux is measured. Since $L$ is a bolometric
luminosity, a frequency dependent correction must be applied to obtain
the monochromatic rest-frame luminosity $L_\nu$ and hence observed
flux $s_\nu(\nu_o)=(1+z)L_\nu(\nu)/4\pi d_L^2$, where $\nu$ is the
rest-frame frequency, $\nu_o=\nu/(1+z)$ is the observer frame
frequency and $d_L$ is the luminosity distance. This correction is
similar to a standard k-correction and depends on the galaxy spectral
energy distribution (SED -- see Section \ref{sec_synth_data}).

Using equations (\ref{eq_lf_evol})
and (\ref{eq_ncounts}), the number of galaxies that contribute to
$n_i$ from a given pixel $j$ in the $L-z$ plane can be written
\be
\label{eq_ncontrib}
m_{ij} \simeq e_j \, \int^{z^j_u}_{z^j_l} {\rm d}V(z) 
\int^{L^{ij}_u}_{L^{ij}_l} {\rm d}L \, \phi_0(L) \, = e_j\,a_{ij} \, .
\ee
Here, the volume integral is written assuming the $j$th pixel
in the $L-z$ plane has fixed redshift boundaries. In general, 
the $L-z$ plane can be divided up into pixels of arbitrary geometry
in which case the integral would extend over the portion of pixel $j$
that contributes to bin $i$. The luminosity limits are
capped by the luminosity range spanned by pixel $j$, 
giving rise to zero contribution if they lie outside the pixel.
Note also that the luminosity limits
have picked up an additional index, reflecting their dependence
on the redshift limits of pixel $j$. The approximation in equation
(\ref{eq_ncontrib}) arises due to the fact that the continuous
evolution implicit in equation (\ref{eq_ncounts}) has been replaced
by the discrete evolution $e_j$ which is assumed constant over the 
entire pixel. The discretised version of equation (\ref{eq_ncounts})
can thus be written:
\be
n_i=\sum_j \, m_{ij} \, ,
\ee
where the sum acts over all pixels in the $L-z$ plane although only a
fraction will typically contribute to number count bin $i$.  

To incorporate redshift data, the procedure is essentially the same as
for number counts. Redshift surveys provide the most direct
measurement of the luminosity function at a specific place in the
$L-z$ plane.  Instead of binning galaxies by flux, and evaluating the
integral in equation (\ref{eq_ncounts}) over $0 \leq z
\leq \infty$, the data are binned by flux {\em and} redshift and the
integral in equation (\ref{eq_ncounts}) extends only over the redshift
range of the bin. Instead of the quantities $a_{ij}$ defined in equation
(\ref{eq_ncontrib}), an analogous set of quantities $b_{ij}$ are
computed.

Including integrated background flux measurements requires a slightly
different approach. Rather than compute the number of galaxies in a
given data bin, the strength and spectral shape of the extragalactic
background radiation in the far-IR/submm waveband puts a constraint on
the integral of the luminosity-weighted GLF over the entire $L-z$
plane. The predicted integrated background flux, $f_\nu$, as measured
at a given observer-frame frequency, $\nu_o$, is the sum of
contributions from all galaxies over the whole $L-z$ plane, i.e.,
\be
f_\nu(\nu_o)=\int^{z=\infty}_{z=0} {\rm d}V(z)
\int^{\infty}_0 {\rm d}L \, \phi(L,z) \, s_\nu(\nu_o) \, .
\ee
As before, the flux $s_\nu(\nu_o)$ is computed allowing for
the SED dependent correction that converts bolometric luminosity
to monochromatic luminosity in the rest frame of the galaxy.
Writing this equation in its discretised form gives
\ba
\label{eq_fnu_discrete}
f_\nu(\nu_{o,i}) &\simeq& \sum_j e_j \, \int^{z^j_u}_{z^j_l} {\rm d}V(z) 
\int^{L^{ij}_u}_{L^{ij}_l} {\rm d}L \, \phi_0(L) \, s_\nu(\nu_{o,i}) \\ \nn
&=& \sum_j e_j\,c_{ij} =f_{\nu,i}\, ,
\ea
for the predicted integrated background flux measurement at the
$i$th observer-frame frequency $\nu_{o,i}$.

From the observed galaxy number counts, $n^o_i$, the observed number
counts binned by redshift, $n^o_{z,i}$ and the measured background
flux, $f^o_{\nu,i}$ (all of which are assumed to be statistically
independent of each other), the $\chi^2$ statistic in terms of the
discretised predicted quantities is
\ba
\label{eq_chisq}
\chi^2 = \sum_{i=1}^{N_1} \frac{(n^o_i-\sum_j e_j\,a_{ij})^2}{\sigma_i^2}+
\sum_{i=N_1+1}^{N_1+N_2} \frac{(n^o_{z,i}-\sum_j e_j\,b_{ij})^2}{\sigma_i^2} 
 \nonumber \\
+\sum_{i=N_1+N_2+1}^{N_T} \frac{(f^o_{\nu,i}-\sum_j e_j\,c_{ij})^2}{\sigma_i^2}
=\sum_{i}\frac{(g_i-\sum_j e_j\,p_{ij})^2}{\sigma_i^2}\, .
\ea
In the above, there are $N_1$ number count bins, $N_2$ number counts
binned by redshift, $N_T$ data points in total including the
background flux measurements and the $\sigma_i$ are the 1$\sigma$
measurement errors.  To simplify, the last term defines the general
quantity $g_i$ as an observed data point and $p_{ij}$ refers to one of
the corresponding quantities $a_{ij}$, $b_{ij}$ or $c_{ij}$. The set
of values $e_j$ that minimise $\chi^2$ are those that satisfy
$\partial \chi^2 / \partial e_k=0$ which gives
\be
\label{eq_sim_eqns}
\sum_i \, g_i \, p_{ik}/\sigma^2_i =
\sum_i \sum_j \, e_j \, p_{ij} \, p_{ik}/\sigma_i^2 \, .
\ee
By writing the vector element
$d_k=\Sigma_i \, g_i \, p_{ik} / \sigma_i^2$ and the matrix element
$M_{kj}=\Sigma_i \, p_{ij} \, p_{ik}/\sigma_i^2$, the solution to
the set of linear equations in (\ref{eq_sim_eqns}) can be written
\be
\label{eq_matrix}
\mathbf{e} = \mathbf{M^{-1}} \, \mathbf{d} \, ,
\ee
where $\mathbf{e}$ is a vector containing the elements $e_j$.

In the presence of noise, the solution given by 
equation (\ref{eq_matrix}) is formally ill-conditioned
and hence must be regularised. This is achieved by
adding an extra term, the regularisation matrix $\mathbf{R}$,
weighted by the regularisation weight, $\lambda$ (see
Section \ref{sec_regularisation}):
\be
\label{eq_matrix_reg}
\mathbf{e}=(\mathbf{M}+\lambda\mathbf{R})^{-1}\mathbf{d} \,.
\ee
The corresponding covariance matrix was derived by
\citet{warren03} for this problem:
\be
\label{eq_evol_errors}
\mathbf{C} = \mathbf{N} - \lambda \mathbf{N}
(\mathbf{N}\mathbf{R})^T \, , \,\,\,
\mathbf{N}=(\mathbf{M}+\lambda\mathbf{R})^{-1}.
\ee
This innocuous equation brings about the method's key advantage over
previous methods since the covariance matrix contains all of the
information required to determine the uncertainty on the evolution
function for a given set of model parameters (e.g., $\lambda$). The
only extra step required to obtain the total error is to propagate the
additional error that arises as a result of the uncertainty on the
model parameters (see next section).

\subsection{Bayesian Evidence}
\label{sec_evidence}

Although regularisation ensures that the solution given by equation
(\ref{eq_matrix_reg}) is well defined, it unfortunately introduces a
new problem.  Regularising a solution reduces the effective number of
degrees of freedom by an amount that can not be satisfactorily
determined. Furthermore, applying the same regularisation weight to
two different models (for example different pixellisations of the
$L-z$ plane) results in a different effective number of degrees of
freedom for each model. This means the minimum $\chi^2$ is biased away
from the most probable solution. More crucially, comparison between
different models cannot be carried out fairly using the $\chi^2$
statistic. Therefore, $\chi^2$ can not be used to determine the most
appropriate pixellisation given the observed data.

One solution to the problem is to simply not regularise. Fortunately,
a better solution can be found by turning to Bayesian inference and
ranking models by their Bayesian evidence instead of $\chi^2$.
\citet{suyu06} derived an expression for the Bayesian evidence,
$\epsilon$, for the linear inversion problem described by equation
(\ref{eq_matrix_reg}). Using the previous notation, this can be
written \ba
\label{eq_evidence}
-2 \,{\rm ln} \, \epsilon &=& 
\chi^2 -{\rm ln} \, \left( {\rm det} [\lambda\mathbf{R}]\right)
+{\rm ln} \, \left( {\rm det} [\mathbf{M}+\lambda\mathbf{R}]\right)
\nonumber \\
& & + \, \lambda\mathbf{e}^T\mathbf{R\,e} +
\sum_{i}\,{\rm ln} (2\pi \sigma_i^2) \, 
\ea
with $\chi^2$ given by equation (\ref{eq_chisq}).  Here, 
the covariance between all pairs of number count bins has been set to
zero \citep[i.e., it is assumed that all observed data points are 
independent  of each other. For covariant data, the more general 
form given by][would be used]{suyu06}.

The evidence is a probability distribution in the model parameters,
allowing different models to be ranked fairly to find the most
probable model. Formally, the evidence should be marginalised over all
parameters and the resulting probability used in the ranking. However,
\citet{suyu06} noted that a reasonable simplification is to
approximate the evidence distribution as a delta function and use the
maximum evidence directly to rank models \citep[see the appendix
of][for more details]{dye08}. We have verified that this approximation
is still valid for our application and hence have adopted it in the
present study.

In our case, the evidence distribution is a function of two model
parameters; the regularisation weight, $\lambda$, and a parameter,
$\rho_{\rm thresh}$, described in the following section that controls
the average size of adaptive pixels in the $L-z$ plane.  Since the
evolution depends on the value of these two parameters, their
uncertainty, as given by the evidence distribution, must be included
in the overall uncertainty on the evolution. For every $L-z$ plane
pixel, we compute this additional uncertainty, $\sigma_e$, using:
\be
\sigma_e^2 = \sum_j \, \epsilon_j  \, \Delta \rho_{\rm thresh}
\, \Delta(\log \lambda)  \, (e_j-<e>)^2
\ee
where the sum acts over all points on the grid spanned by $\lambda$
and $\rho_{\rm thresh}$ at which the evidence, $\epsilon_j$, is
evaluated. Here, $\Delta \rho_{\rm thresh}$ and $\Delta(\log \lambda)$
is the separation between grid points along the $\rho_{\rm thresh}$
and $\log \lambda$ directions respectively and $<e>$ is the mean
evolution in the pixel. Note also that the evidence must be
normalised such that $\sum_j \, \epsilon_j \, \Delta \rho_{\rm thresh}
\, \Delta(\log \lambda) = 1$. This additional error is therefore
effectively the scatter in the evolution, weighted by the evidence. We
compute the overall error in the evolution as the quadrature sum of
this error and the uncertainty given by the appropriate term in the
covariance matrix given in equation (\ref{eq_evol_errors}).

\subsection{Adaptive gridding of the $L-z$ plane}
\label{sec_adaptive_gridding}

As we alluded to earlier, the $L-z$ plane can be pixellised in a
completely general way. This has the advantage that smaller pixels can
be placed in regions where there are superior observational data,
i.e., higher signal-to-noise (S/N) and/or a higher density of data
points. The effect of this is to better resolve the evolution function
whilst maintaining an approximately constant level of constraints per
pixel over the $L-z$ plane. Note that although the method prescribed
by Peacock \& Gull can adjust to the data by limiting the series
expansion of the evolution function, this alters the {\em global}
resolution over the plane, not in specific regions that are better
constrained.

There are several criteria that could be used to control the size of
pixels in the $L-z$ plane. One example would be to make direct use of
the density of data points weighted by their S/N. Our criterion of
choice is to use the covariance between pixel pairs provided by
equation (\ref{eq_evol_errors}).  Joining two pixels together
increases the statistical independence and hence lowers the covariance
of the resulting larger pixel with its neighbours. Defining a
covariance threshold, $\rho_{\rm thresh}$, therefore introduces a
criterion whereby pixel pairs whose covariance is larger than
$\rho_{\rm thresh}$ are joined together.

The procedure we have used for adaptively gridding the $L-z$ plane 
is as follows:
\begin{itemize}

\item[1)] The $L-z$ plane is uniformly pixellised with a regular
grid of small rectangular pixels. We tested a range of initial grid
sizes and found that a $20 \times 20$ grid is a good compromise
between having sufficient resolution to properly adapt to the varying
range of covariances over the $L-z$ plane and being of low enough
resolution to maintain a high execution speed.

\item[2)] For a given regularisation weight, the covariance matrix
given by equation (\ref{eq_evol_errors}) is computed.

\item[3)] The lower triangle of the covariance matrix is scanned for
elements with an absolute value larger than the threshold
covariance. Working from largest to smallest, pixel pairs whose
covariance exceeds the threshold are joined. If, in working down the
list, a pixel pair is encountered where one of the pixels has already
been joined (because it had a higher covariance with another pixel),
that pixel pair is skipped.

\item[4)] New versions of the matrices $\mathbf{M}$ and $\mathbf{R}$
are computed taking into account the new pixellisation.

\end{itemize}
Steps 2) to 4) are repeated until the absolute value of all pixel
covariances lie beneath the threshold $\rho_{\rm thresh}$.  A more
formally correct procedure would be to recompute the covariance matrix
every time a pair of pixels are joined.  This is considerably slower
to execute in practice and we have found that the resulting adapted
grid does not differ significantly from that produced by the faster
procedure outlined above.

\subsection{Regularisation}
\label{sec_regularisation}

The regularisation matrix introduced in equation (\ref{eq_matrix_reg})
arises as a result of adding a term to $\chi^2$ that becomes
numerically smaller for smoother solutions.  In this sense,
regularisation behaves as a prior, acting to penalise noisy evolution
functions. This is necessary to prevent ill-conditioning in equation
(\ref{eq_matrix}). A potential disadvantage is that the evolution may
not be smooth in reality so that regularisation biases the solution.
However, the regularisation weight is set by the evidence which is
maximised at the optimal value of the regularisation weight $\lambda$.
If more data points are available, or if the data have a higher S/N,
the evidence is maximised at a lower value of $\lambda$, allowing
greater resolution of any sharp features in the reconstructed
evolution.  We have investigated this to an extent with a synthetic
dataset produced using an evolution function with a cutoff (see
Section \ref{sec_demo}).

The regularisation term added to $\chi^2$ can be written
\be
\label{eq_reg_term}
\lambda B=\lambda \sum_{j,k}\,R_{jk}e_j\,e_k
\ee
where the $R_{jk}$ are the elements of the regularisation matrix
$\mathbf{R}$. These depend on the regularisation
scheme adopted as described below. This means that the solution 
remains linear since the partial derivative of $B$ with respect 
to the the $e_j$ is linear in $e$. 

The simplest form of regularisation is {\em zeroth order}
regularisation, where solutions that deviate from zero are penalised
using $w_{jk}=\delta_{jk}$ so that $B=\sum_j\,e_j^2$. More useful
forms of regularisation are {\em first order} schemes which penalise
solutions that deviate from a constant and {\em second order} which
penalise solutions that deviate from a gradient. 

For most solutions, as we have found in carrying out the work
presented herein, first and higher order regularisations give very
similar results. Whilst we could implement first order regularisation,
higher orders are generally more effective in terms of allowing more
freedom in the solution. However, conventional higher order schemes
are ill-defined on our adaptively gridded $L-z$ plane.  For example,
with a second order gradient scheme, pixel triplets required to form a
finite difference second order derivative would be neither co-linear
nor parallel. With this in mind, in this work, we have opted for a
hybrid scheme that uses the difference between the evolution in a
given pixel $j$ and the sum of evolutions in all neighbouring pixels
$k$ weighted by
\be
\label{eq_reg_wts}
h_{jk}=(\Omega_k/\Omega_j)\,N\exp(-y_{jk}^2/r^2) \, .
\ee
Here, $\Omega$ is the area of the pixel in the $L-z$ plane, $y_{jk}$
is the separation of the centres of pixels $j$ and $k$ and $N$ is a
normalisation constant set such that $\sum_{k,k\ne j} h_{jk}=1$ and
$N=1$ when $j=k$. We set $r=1$ in units of pixels.  The matrix
$\mathbf{h}$ composed of elements given by equation (\ref{eq_reg_wts})
relates to the regularisation matrix $\mathbf{R}$
via $\mathbf{R}=\mathbf{h}^{T}\mathbf{h}$. By construction,
$\mathbf{R}$ is non-singular so that the evidence given by
equation (\ref{eq_evidence}) is always calculable.

As a final note, all regularisation weights quoted in this paper are
scaled by the ratio of traces Tr$[\mathbf{M}$]/Tr$[\mathbf{R}]$.
This normalisation factor means that the regularisation term 
$\lambda\mathbf{e}^T\mathbf{R\,e}$ and $\chi^2$ in equation
(\ref{eq_evidence}) are weighted approximately equally when
the scaled regularisation weight is unity.

\subsection{Evidence maximisation procedure}

There are two non-linear variables which must be adjusted when
searching for the maximum evidence. These are the regularisation
weight, $\lambda$ and the covariance threshold, $\rho_{\rm thresh}$.
We also experimented with allowing the initial grid resolution to vary
as a free parameter but found a strong degeneracy with $\rho_{\rm
thresh}$ (if smaller pixels are used, more are joined to give a
similar final adaptive grid).

With every trial set of $\lambda$ and $\rho_{\rm thresh}$, the
procedure we have adhered to is:
\begin{itemize}

\item[1)] Form an adaptive grid (see Section \ref{sec_adaptive_gridding}).

\item[2)] Form the regularisation matrix $\mathbf{R}$ for the current
adaptive grid.

\item[3)] Compute the matrix $\mathbf{M}$ and vector $\mathbf{d}$ as
described in Section \ref{sec_method}. From these, solve for
the evolution as given by equation (\ref{eq_matrix_reg}). We use
an $LU$ factorisation routine with iterative refinement.

\item[4)] Compute the covariance matrix as given by equation
(\ref{eq_evol_errors}), $\chi^2$ given by equation (\ref{eq_chisq})
and then ${\rm ln} \, \epsilon$ from equation (\ref{eq_evidence}).

\end{itemize}

To find the maximum evidence, we performed a grid search over
regularly stepped values of log$\lambda$ and $\rho_{\rm thresh}$. The
nature of the adaptive gridding process is such that the evidence
surface is not perfectly smooth. Despite smoothly varying $\lambda$
and $\rho_{\rm thresh}$, discontinuous jumps in the evidence occur
when pixels are joined together. By choosing a smaller pixel grid, the
size of the discontinuities are reduced. Our findings indicate that,
provided the numerical evaluation of the integrals in equations
(\ref{eq_ncontrib}) and (\ref{eq_fnu_discrete}) is precise, the
evidence surface is sufficiently smooth with our chosen $20 \times 20$
grid for easy identification the global maximum. In light of this, a
more efficient automated search for the maximum such as a downhill
simplex method could be used.  In the results presented in the next
section, we show contour plots of the evidence as a function of
log$\lambda$ and $\rho_{\rm thresh}$.

\section{Application to synthetic datasets}
\label{sec_demo}

In this section, we demonstrate the method by applying it to synthetic
datasets. These datasets are created from an input evolution function
which can be compared with the reconstructed evolution functions.

\subsection{Synthetic dataset construction}
\label{sec_synth_data}

\begin{figure}
\epsfxsize=80mm
{\hfill
\epsfbox{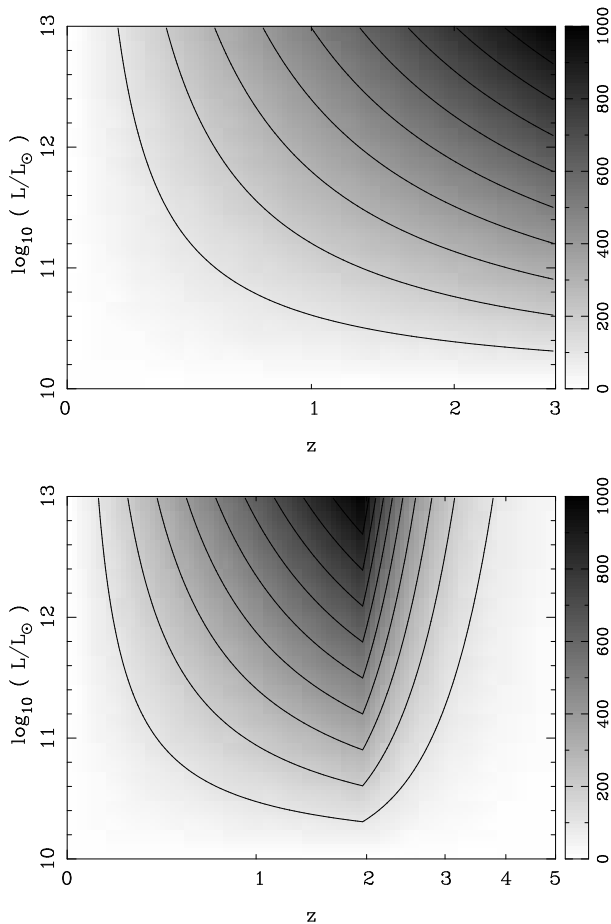}
\hfill}
\caption{The two input evolution functions. {\em Top}: The monotonic
function which increases with increasing redshift and luminosity.
{\em Bottom}: The cutoff function formed by multiplying the monotonic
function by an exponential cutoff $\propto e^{2-z}$ for $z\geq 2$. 
Contours in both plots are in intervals of 100.}
\label{input_evol_fns}
\end{figure}

Three ingredients are required to synthesize a dataset:
\begin{itemize}

\item[1)] {\em Dataset characteristics}: In the case of number
counts, these are the flux limits, in the case of number
counts binned by redshift, the flux and redshift limits and
for background flux measurements, the frequencies at which the
flux is measured. In addition, all three types of dataset 
must stipulate a survey area for determination of uncertainties.

\item[2)] {\em The local luminosity function}: To keep our
demonstration reasonably realistic, we have used a luminosity function
derived from local galaxies (selected with velocities $<30,000
\,$km/s) in the Infra-Red Astronomical Satellite (IRAS) Point Source
Catalogue of \citet{saunders00}. In order to calculate the luminosity
of each galaxy at each frequency of interest and to compute the
bolometric luminosity, we estimated an SED. Each SED was fit to the 60
and 100$\,\mu$m IRAS flux and an estimate of the 850$\,\mu$m flux
provided by the tight correlation between 60, 100 and 850$\,\mu$m flux
reported by \citet{vlahakis05}. We refer the reader to a forthcoming
study where we apply the method to real data (Dye et al., in
preparation) for more details.

\item[3)] {\em The evolution function}: We used two different
evolution functions. The first is motivated by the findings of
\citet{heavens04} and is a smooth function that monotonically
increases with increasing redshift and luminosity. We used a bi-linear
interpolation in $\log (1+z), \log L$ space according to
$E(L,z)=1+xyE_{max}$ where $x=\logten (1+z)/\logten (1+z_{max})$,
$y=(\logten L)/\logten (L_{max}/L_{min})$ and $E_{max}=1000$ is the
maximum evolution at the point $(L_{max},z_{max})$ in the $L-z$
plane. The second model evolution function is the product of the first
and an exponential cutoff $1.66\,e^{2-z}$ which applies at $z\geq 2$.
The factor 1.66 ensures that the maximum evolution at $z=2$ is equal
to 1000. The monotonic evolution function is defined over $0 \leq z
\leq 3$ whereas the cutoff function extends to $0 \leq z \leq 5$.
Both are defined over $10^{10} \leq L/L_\odot \leq 10^{13}$. Figure
\ref{input_evol_fns} plots the functions.

\end{itemize}

In the case of number counts and number counts binned by redshift, we
randomised the synthetic data according to Poisson
statistics. Datasets were generated assuming either 0.1 deg$^2$
coverage (the 'low S/N' case) or 10 deg$^2$ coverage (the 'high S/N'
case).  Since background flux estimates are measured via a different
means in practice, we have randomized the multi-frequency set of
synthetic background fluxes with a 10\% Gaussian error.

For all four combinations of S/N and evolution function, we have 
generated four different datasets:
\begin{itemize}

\item[A)] Number counts at 850$\,\mu$m between the limits 2$\,$mJy
(the confusion limit at 850$\mu$m with a 15$\,$m dish)
and 200$\,$mJy in six flux bins equally spaced in log flux.

\item[B)] Number counts at 850$\,\mu$m between the limits
0.1 and 200$\,$mJy in 10 flux bins equally spaced in log flux.

\item[C)] Number counts at 850$\,\mu$m between the limits
0.1 and 200$\,$mJy in 10 intervals binned by 10 redshift intervals
equally spaced in log($1+z$).

\item[D)] A set of 10 background fluxes evaluated over the frequency
range 2.4$\,$THz - 300$\,$GHz (125$\,\mu$m - 1$\,$mm) at regularly
spaced intervals in log(frequency).

\end{itemize}

Given currently available observing facilities, dataset C is somewhat
unrealistic. Redshift measurements down to fluxes of $\sim 0.1
\,\mu$mJy at 850$\,\mu$m are well below the confusion limit of 15m
single dish observations and higher resolution interferometry
presently lacks the sensitivity to generate large source samples.
However, our choice to include this dataset is motivated by the fact
that the dataset provides an indication of the best possible
performance that can be expected from the method, thereby setting a
benchmark for future surveys to aspire to.

The next section describes application of the reconstruction method to
different combinations of these datasets. In addition to this, Section
\ref{sec_future_surveys} outlines further tests of the method with a
more realistic set of data based on what forthcoming {\sl Herschel}
and SCUBA2 surveys are expected to deliver.

\subsection{Evolution reconstructions}
\label{sec_results}

We have applied our method to various dataset combinations to test how
its performance depends on data availability and quality.  The first
and second groups of reconstructions in Sections \ref{sec_mono_recon}
and \ref{sec_cutoff_recon} apply to datasets generated using the
monotonic and cut-off evolution function respectively (see Figure
\ref{input_evol_fns}). In the third group presented in Section
\ref{sec_high_sn_recon}, we compare recovery of the evolution with
high S/N versions of datasets in the first and second groups. Finally,
the last set of reconstructions in Section \ref{sec_future_surveys}
are based on synthetic {\sl Herschel} and SCUBA2 datasets.

\subsubsection{Monotonic evolution}
\label{sec_mono_recon}

\begin{figure*}
\epsfxsize=180mm
{\hfill
\epsfbox{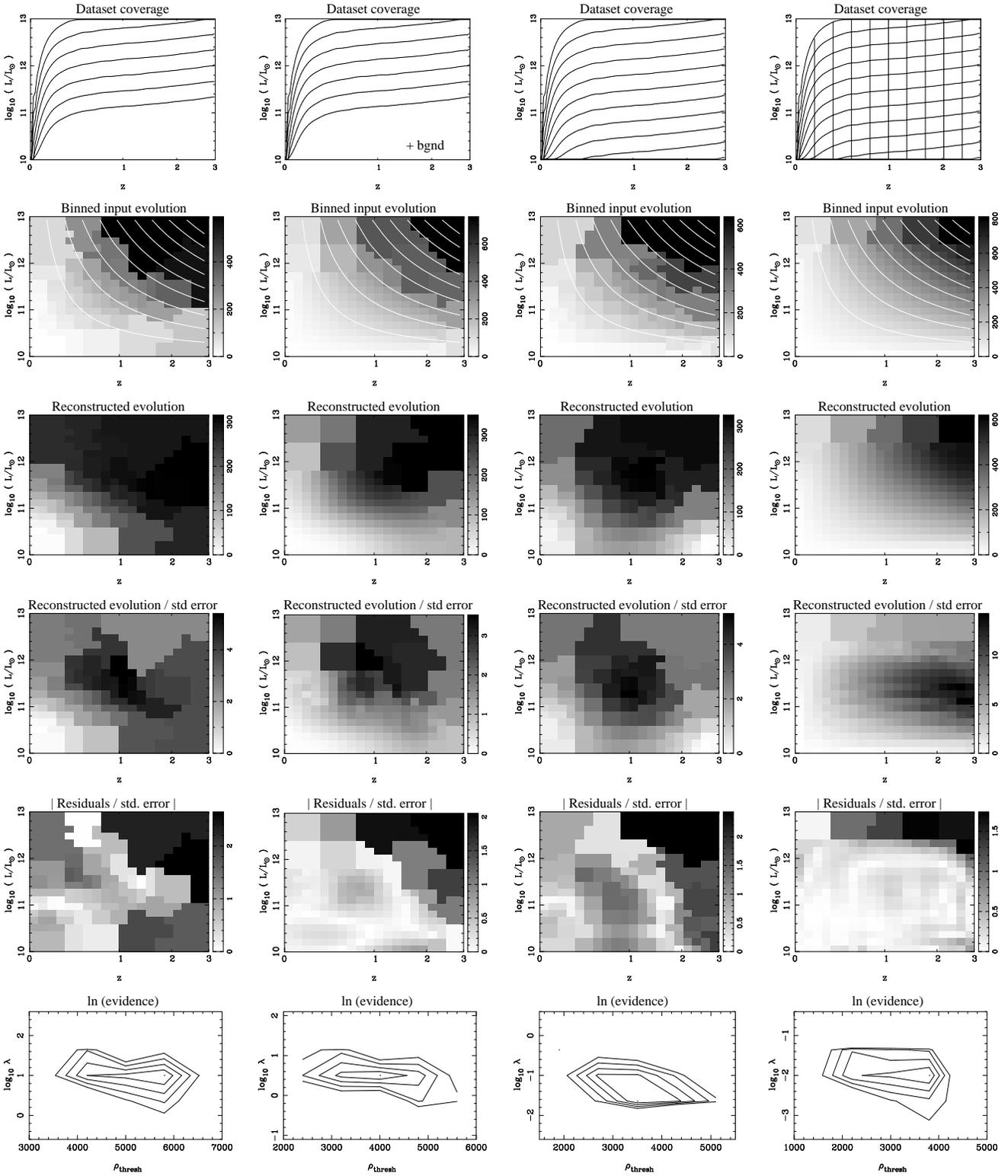}
\hfill}
\caption{Application of the reconstruction method to four different
combinations of low S/N datasets generated using the monotonic
evolution function. Reading from left to right, columns correspond to
datasets A, A+D, B, C (see Section \ref{sec_synth_data}). Reading from
top to bottom, each row shows 1) coverage of input dataset(s), 2)
input evolution function binned to the adaptive grid determined in the
reconstruction (contours show the smooth underlying function plotted
in Figure \ref{input_evol_fns}), 3) reconstructed evolution function,
4) the significance of the reconstructed evolution function, 5) the
significance of the absolute residuals between the binned input and
reconstructed evolution functions, 6) evidence contours in the plane
spanned by covariance threshold and regularisation weight. Contours
are stepped in four intervals of $\Delta \ln \epsilon=1$ away from the
maximum.}
\label{recon1}
\end{figure*}

The first group of reconstructions are shown in Figure
\ref{recon1}. These apply to the four dataset combinations A, A+D, B
and C (see Section \ref{sec_synth_data}) generated using the monotonic
evolution function in the low S/N case. The top row in the figure
plots the coverage of each dataset in the $L-z$ plane.  For each of
the four test reconstructions shown, we have binned the input
evolution function to the adaptive grid determined in each case to
facilitate comparison with the recovered evolution. The binned input
and recovered evolution is plotted in the second and third rows of the
figure respectively. Comparing the two clearly shows that recovery of
the input evolution becomes increasingly more faithful as the coverage
of the datasets in the $L-z$ plane increases. The average size of
adaptive pixels selected by the maximum evidence decreases with
increasing coverage as a result of the increased constraining power of
the data. Similarly, as the evidence contour plots at the bottom of
the figure show, less regularisation of the evolution function is
necessary when the data coverage is more extensive. Notice also that
the covariance threshold at which the evidence is maximised is lower
when the constraints provided by the data are greater. Finally, an
important point to note is that in all cases, the significance of the
residuals is distributed approximately normally with unit standard
deviation. This means that the uncertainty on the evolution derived by
our procedure is an accurate representation of where the recovered
evolution truly differs from the input evolution.

The evidence contours show that a mild degeneracy exists between the
covariance threshold and the regularisation weight. The degeneracy
arises because a higher regularisation weight imposes stronger
constraints on pixels. This has the effect of reducing their average
covariance as quantified by equation
(\ref{eq_evol_errors}). Therefore, at a fixed covariance threshold, as
the regularisation weight increases, the average pixel size decreases.
To maintain an optimal adaptive grid (i.e., an optimal average pixel
size) at higher regularisation weights, a lower covariance threshold
must be used and vice versa. In fact, the inclination of the evidence
contours is parallel to lines of constant numbers of pixels. The
maximum evidence therefore selects a specific subset of adaptive grids
with approximately the same number of pixels from the set of all
possible grids. The contours do not extend indefinitely because at the
low covariance threshold end, the higher regularisation reduces the
dynamic range of covariances over the $L-z$ plane. This results in an
adaptive grid with more similarly sized pixels, i.e., the grid is not
able to optimally adapt to the variation in data coverage. The reverse
is true at the high covariance threshold end, where the lower
regularisation results in a larger range of covariances to the extent
that some pixels become too large while some remain too small to
properly adapt to the data coverage.

Considering each reconstruction in turn in Figure \ref{recon1}, it is
apparent that number count data alone are unable to recover the input
evolution to a satisfactory accuracy. In the case of dataset A shown
in the first column (with number counts that reach down to 2$\,$mJy
and no background flux measurements), the evolution is biased low at
high $L$, high $z$ and biased high at high $L$, low $z$ and at low
$L$, high $z$. This biasing occurs because the evolution function has
been 'smeared' over the $L-z$ plane by two effects. The first effect
causes the evolution to be smeared along the lengths of flux
bins. This is a direct result of the fact that the number of observed
galaxies in a given bin can be reproduced by many different evolution
profiles along the bin's length, a point we will return to below. The
second effect is that the regularisation effectively smooths the
evolution. Regions of the $L-z$ plane where there are no constraints
from the data are only constrained by the regularisation. In these
regions, the evolution is extrapolated from neighbouring
data-constrained regions, by an amount governed by the regularisation
weight. This is a useful feature which we alluded to in the
introduction: The ability to extrapolate is one of the key advantages
of model-based approaches to measuring evolution.

It is apparent that the residuals in the first column of Figure
\ref{recon1} show that the most significant deviation from the input
evolution occurs not where data constraints are lacking, but at high
$L$, high $z$ where the evolution is strongest. This is again due to
the smearing which causes a larger absolute difference when the
evolution is stronger. In this sense, it could be argued that the
method has performed more reliably in regions of the $L-z$ plane not
covered by data.  Of course this is primarily because the method
assigns a much higher uncertainty to areas of zero coverage. The ratio
of input to recovered evolution in areas of zero data coverage shows a
stronger deviation from unity than areas that are covered by data.

Despite the biasing, the number count data still successfully detect a
significant increase in the evolution from low $L$, low $z$ to high
$L$, high $z$. At first sight, this seems incredible given that number
count data contain nothing more than the number of galaxies between
two flux limits. In other words, a single flux bin gives only a
measure of the average evolution within it and nothing about the
variation of the evolution. The same measured number of galaxies
between two flux limits can be reproduced with a wide range of
different evolution functions. Although true of an individual bin,
this degeneracy becomes greatly reduced when other flux bins
are added to the constraining data, especially when the solution is
regularised. 

\begin{figure}
\epsfxsize=85mm
{\hfill
\epsfbox{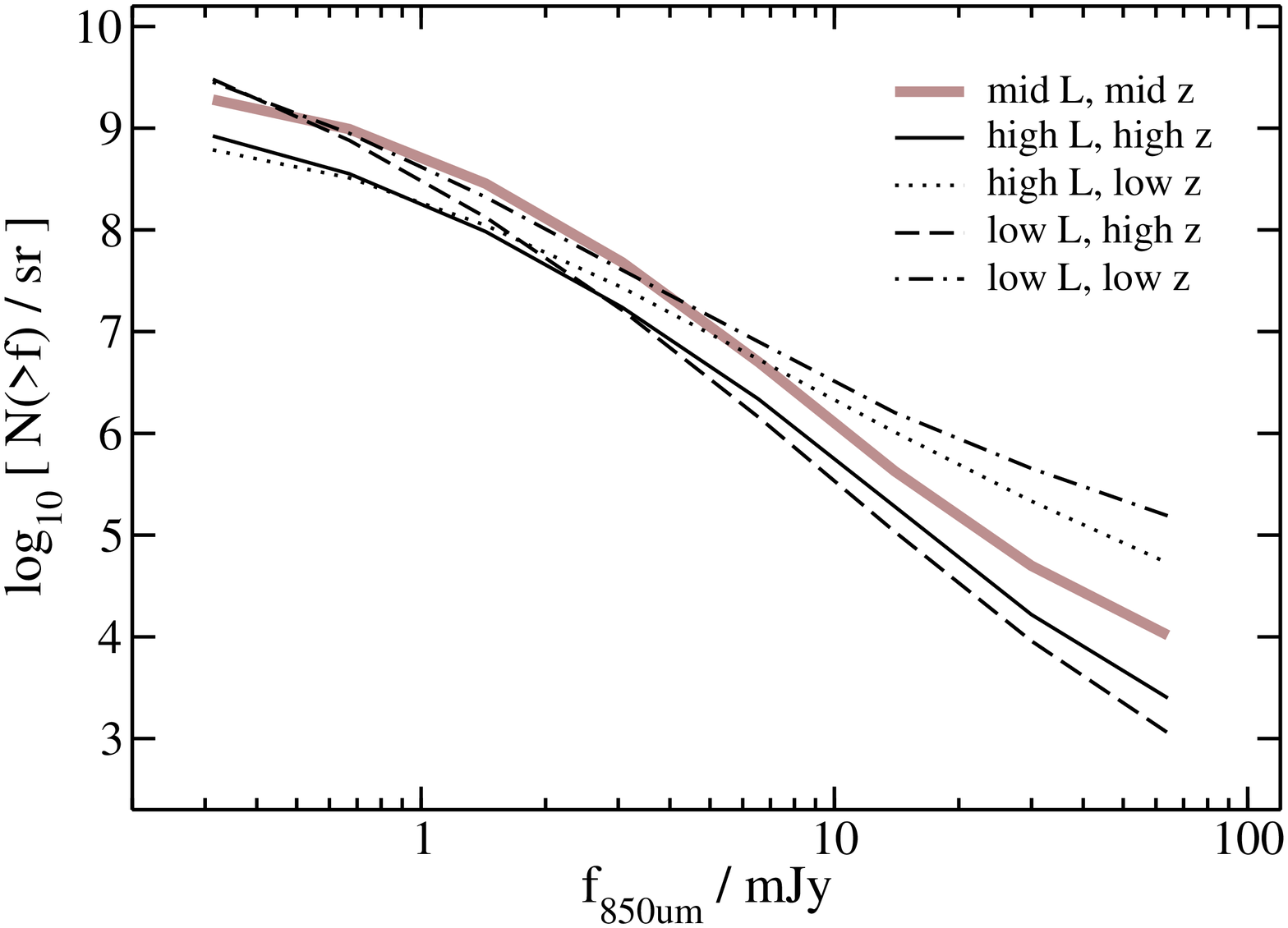}
\hfill}
\caption{Synthetic 850$\,\mu$m cumulative number counts generated with
five different evolution functions. Four of these are the four
90-degree rotations of the monotonic evolution function plotted in
Figure \ref{input_evol_fns} such that the maximum lies at each of the
four corners of the $L-z$ plane. The fifth evolution function is
described by a Gaussian with a full-width-at-half-max of $\Delta \log
L = 1$ and $\Delta z = 1$, reaching the same maximum evolution but
peaking in the centre of the $L-z$ plane. The counts arising from this
fifth evolution function are shown by the broader grey line.}
\label{number_counts}
\end{figure}

Figure \ref{number_counts} illustrates the constraining power of pure
number count data. We have plotted synthetic 850$\,\mu$m number counts
generated using five different evolution functions covering $10^{10}
\leq L/L_\odot \leq 10^{13}$ and $0 \leq z \leq 3$ as in Figure
\ref{recon1}. Four of these are monotonic and are simply the four
90-degree rotations of the previous monotonic function so that the
maximum evolution is located at each of the four corners of the $L-z$
plane. The fifth evolution function was chosen to be a Gaussian with a
full-width-at-half-max of $\Delta \log L = 1$ and $\Delta z = 1$,
reaching the same maximum evolution but peaking in the centre of the
$L-z$ plane.

The figure shows that the four different monotonic evolution functions
give rise to four unique number count datasets.  All four
datasets remain unique if any one undergoes the renormalisation $n(f)
\rightarrow \alpha n(f)$ where $\alpha$ is a scale factor. This means
that none of the four evolution functions can be renormalised to
create a set of number counts that match those obtained using any of
the other three evolution functions.  Reversing this argument, the
number count data can therefore distinguish between the four evolution
functions. However, the Gaussian evolution function produces number
counts which are very similar under renormalisation to the monotonic
function that peaks at high $L$, high $z$ (compare the broader grey
line with the thin continuous black line in Figure
\ref{number_counts}). This occurs because translating the Gaussian
function by $\Delta z = +1.5$ along the direction of the 850$\,\mu$m
flux bins (i.e., so that the peak lies at $z=3$), results in an
evolution function that is very similar to the high $L$, high $z$
monotonic function. This serves to demonstrate the insensitivity of
flux bins to variations in evolution along their length. The four
rotated monotonic functions can be distinguished because they have
moments perpendicular to lines of constant flux in the $L-z$
plane. Indeed, the reconstructed evolution obtained from the number
counts generated using the Gaussian evolution function is very similar
to the monotonic function which peaks at high $L$, high $z$, apart
from a normalisation offset. Number counts can therefore not directly
measure evolution cut-offs in redshift. We return to this point below.

The second column in Figure \ref{recon1} shows the method applied to
the combination of datasets A and D (i.e., 850$\,\mu$m number counts
down to 2$\,$mJy and 10 measurements of the background). Compared with
just the number counts shown in the first column, it is clear that the
reconstruction is more accurate. The added constraints from the
background flux reduce the level of smearing. This is particularly
apparent at low luminosity and high redshift where the biasing is much
smaller. The additional constraints also enable a higher resolution
reconstruction at low $L$, low $z$ and result in a lower
required level of regularisation.

In the third column of Figure \ref{recon1}, we show how well the
evolution is recovered with dataset B which comprises pure number
count data again, but down to a lower flux limit of 0.1$\,$mJy. In
this case, the flux bins cover almost the entire $L-z$ plane. The
figure shows that a similar, if not slightly higher level of biasing
in the recovered evolution occurs, compared to that seen in the second
column, due to the same smearing effects.  However, there is an
obvious improvement compared to the shallower number counts shown in
the first column. This is apparent at low $L$ where the shallow counts
permit a lower resolution reconstruction due to lack of coverage.
Notice also that the solution requires a much lower level of
regularisation.

A point to note is that the reconstructed evidence in regions of the
$L-z$ plane constrained by data are not noticeably biased by regions
that are not constrained by data. For instance, comparing the first
and third columns of Figure \ref{recon1} in the region of the $L-z$
plane covered by the brightest six flux bins shows that the recovered
evolution and the significance of the residuals are very similar.
Therefore, interpolation into areas lacking data coverage can be
relied upon without detriment to the overall solution.

The fourth column of Figure \ref{recon1} corresponds to the case where
redshift information is included.  Not surprisingly, the addition of
redshifts improves the reconstruction dramatically by providing direct
measurements of the evolution at specific points in the $L-z$
plane. The dataset allows reconstruction of the evolution on a much
higher resolution grid. The larger pixels that still remain at high
luminosities are due to increased Poisson noise caused by a
significantly lower number density of galaxies (a common feature of
most of the results shown in this paper). Despite the increased
resolution, the significance of the reconstructed evolution is $\sim
3$ times higher compared to the cases where redshifts are not
present. The solution is again biased most at high $L$, high $z$
where the evolution peaks and smearing has the largest effect, 
but this is less extensive and much reduced in magnitude.

\subsubsection{Evolution with a cut-off}
\label{sec_cutoff_recon}

\begin{figure*}
\epsfxsize=180mm
{\hfill
\epsfbox{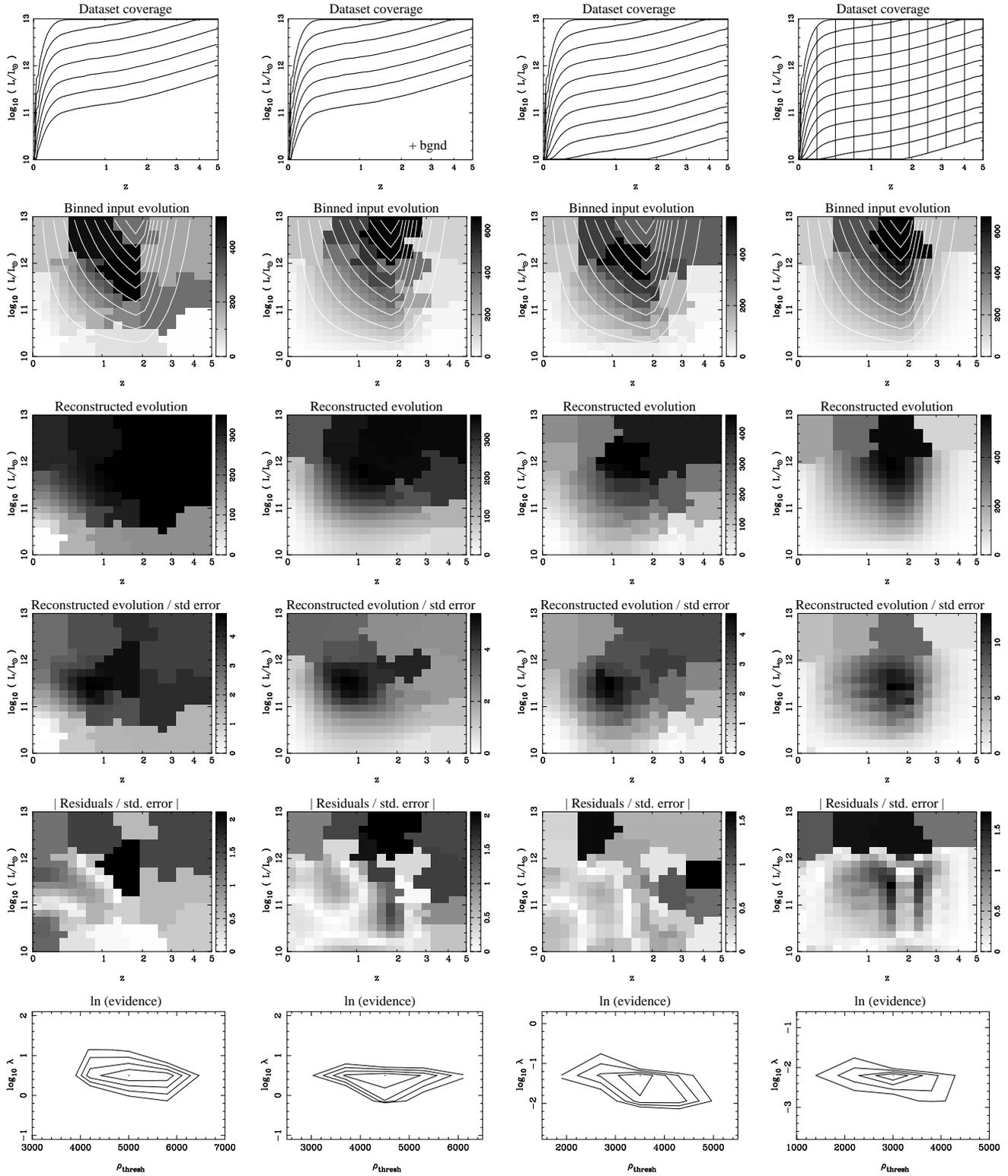}
\hfill}
\caption{As Figure \ref{recon1} but applied to low S/N datasets
generated using the cut-off evolution function.}
\label{recon2}
\end{figure*}

The second group of reconstructions apply to the four dataset
combinations A, A+D, B and C (see Section \ref{sec_synth_data})
generated in this case using the cut-off evolution function with low
S/N (i.e. assuming an area of 0.1$\,$deg$^2$). The top row in Figure
\ref{recon2} shows the coverage of the datasets over the $L-z$ plane
which this time extends further in redshift out to $z=5$.

In terms of the variation of characteristics such as resolution,
residuals, significance of the reconstructed evolution, optimal
regularisation weights and optimal covariance thresholds, the
reconstructions exhibit very similar traits to those seen with the
previous group of datasets generated using the monotonic evolution
function. However, there is one obvious and significant
difference. Although the monotonic evolution function appears to be
fairly well reconstructed without redshift information, the cut-off
evolution function can not be properly recovered unless redshifts are
included. Smearing along the length of flux bins removes any sign of
the cut-off, pushing the strongest evolution to the highest
redshift. This is the exact same behaviour observed previously with
the ad-hoc Gaussian evolution function. We thus iterate again that
redshift data is essential to measure any breaks or discontinuous
features in the evolution.

We have investigated where, in relation to the break, redshift data
most effectively constrains the break. Our findings show that redshift
measurements lower than the break make little or no difference to
identifying the location or even the existence of a break. Only
redshift measurements at the break and beyond are able to constrain
its location.

\subsubsection{High S/N reconstructions}
\label{sec_high_sn_recon}

\begin{figure*}
\epsfxsize=180mm
{\hfill
\epsfbox{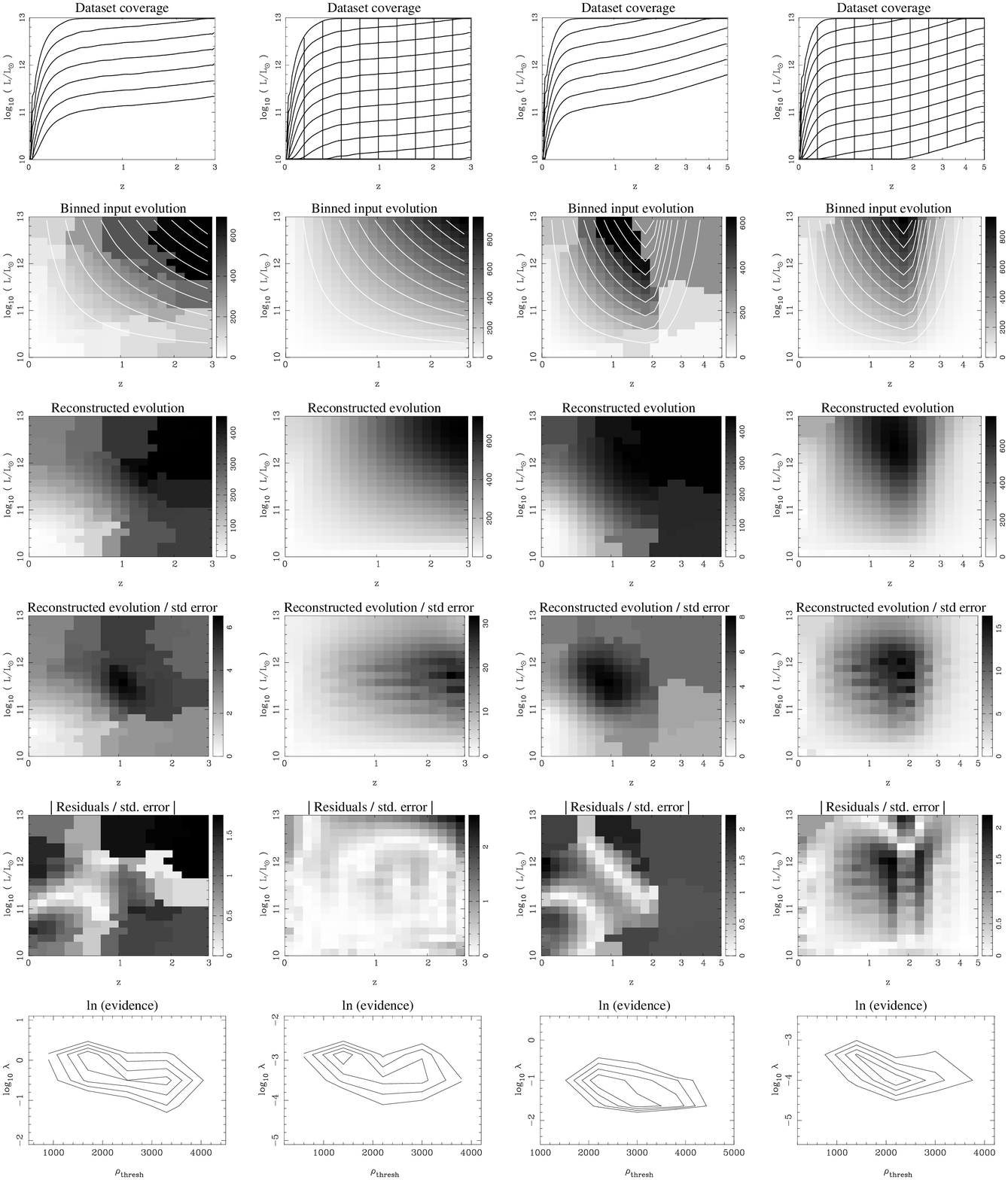}
\hfill}
\caption{Application of the reconstruction method to high S/N
datasets. The first and second columns correspond respectively to
datasets A and C generated with the monotonic evolution function and
the third and fourth columns to datasets A and C generated with the
cut-off evolution. Row descriptions are the same as those in
Figure \ref{recon1}.}
\label{recon3}
\end{figure*}

We have repeated some of the reconstructions carried out previously
but with high S/N datasets (i.e. where an area of 10$\,$deg$^2$ has
been assumed). We generated the two extremes in datasets considered,
namely the shallow number count only dataset and the deeper number
counts binned by redshift, using both the monotonic and the
cut-off evolution function. Figure \ref{recon3} shows the results.

Comparing with the corresponding low S/N reconstructions, the most
striking difference is the improved resolution and higher significance
of the reconstructed evolution. The rib-like structures present in the
evolution significance (also seen to a lesser degree in the low S/N
case) occur because of the varying number of constraints per pixel
which range from one to four data bins depending on their alignment
with the $L-z$ plane pixel grid.

An increase in the accuracy (i.e., a reduction in the average absolute
size of the residuals) is also clear. Although the range of residual
significances is approximately the same between high and low S/N
datasets, the extent of the most significant discrepancies is less in
the high S/N cases.  In fact, the distribution of residual
significances is slightly narrower than a normal distribution with
unit $1 \sigma$ width, implying that the errors are probably slightly
over-estimated. If the residuals were randomly scattered over the
$L-z$ plane, then this would imply that the evolution had been
recovered to the best accuracy possible.  However, the fact that the
residuals form coherent patterns resembling the evolution function
indicates that the reconstruction is not perfect. There are
still low-level biases present, an inescapable consequence of
regularisation.

\subsubsection{Forthcoming survey predictions}
\label{sec_future_surveys}

The reconstructions previously considered give a good indication of
how well different dataset types are able to constrain the
evolution. From these simulations, it is clear that redshift data is
vital to properly recover the input evolution. However, as we
discussed in Section \ref{sec_synth_data}, the extent of the redshift
data coverage in the $L-z$ plane was overly optimistic. Furthermore,
in practice, observational data gathered for the purpose of measuring
evolution will most likely be heterogeneous, comprising redshift and
number count measurements from different studies observed at different
wavelengths and to different depths.

In this section, we investigate the performance of the method with
synthetic data generated to mimic a more realistic collection of data,
typical of what might be expected from forthcoming submm surveys. Two
of the largest datasets will be delivered by {\sl Herschel} and
SCUBA2. In the case of {\sl Herschel}, we synthesized data based on
the Spectral and Photometric Imaging Receiver (SPIRE) which operates
in three passbands with central wavelengths\footnote{We have not
considered the two shorter wavelength passbands of the Photodetector
Array Camera and Spectrometer on board {\sl Herschel} since the
corresponding flux bins only cover a small region of our $L-z$ plane.}
250, 350 and 500$\,\mu$m.  For SCUBA2, we synthesized data at its
longer wavelength channel at 850$\,\mu$m.

In terms of redshifts, we have assumed that measurements can be made
down to the flux confusion limit at each wavelength. For SPIRE, these
limits, as estimated by \citet{lagache03}, are approximately 10, 20
and 20$\,$mJy at 250, 350 and 500$\,\mu$m respectively. The confusion
limit for SCUBA2 at 850$\,\mu$m, is $\sim 2\,$mJy
\citep[e.g.,][]{hughes98}. In terms of number counts, we have assumed
that a flux limit $20\times$ lower than the confusion limit can be
achieved using techniques such as the so-called 'P(D)' statistical
method \citep[see, for e.g.,][and references therein]{patanchon09} or
to a lesser extent, exploitation of gravitational lens magnification
(e.g., Smail, Ivison \& Blain 1997 or more recently, Knudsen et al.
2006).

\begin{figure}
\epsfxsize=90mm
{\hfill
\epsfbox{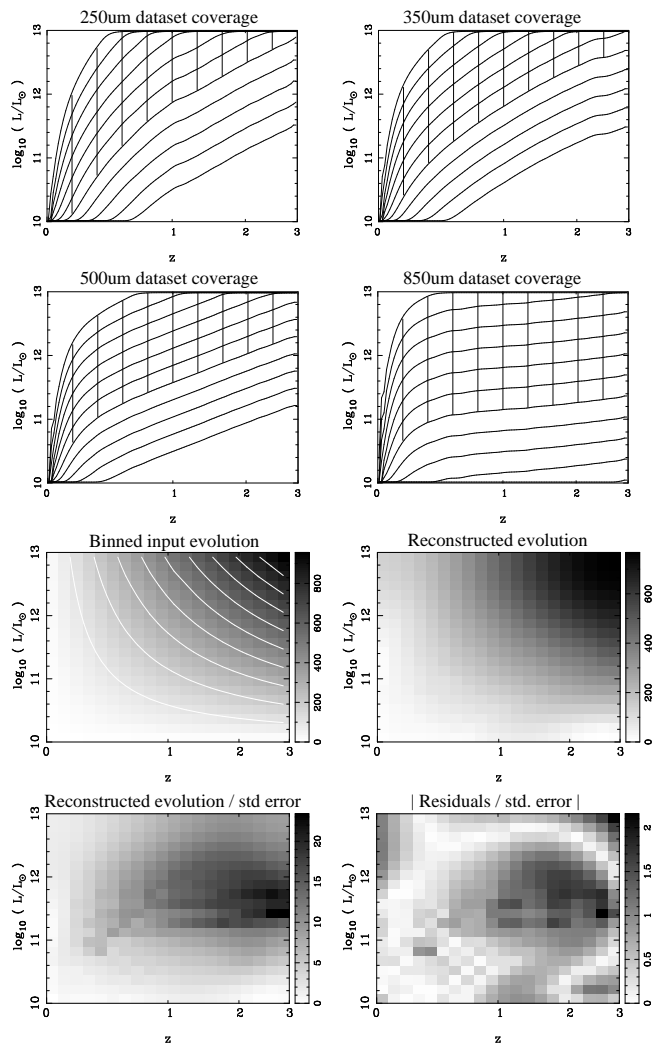}
\hfill}
\caption{Application of the reconstruction method to simulated {\sl
Herschel}/SPIRE (250, 350 and 500$\,\mu$m) and SCUBA2 (850$\,\mu$m)
data generated using the monotonic evolution function over
1$\,$deg$^2$. The top four panels show the coverage of each dataset.
In each case, redshift data extends down to the confusion limit and
number counts reach down another factor of 20 in flux (see main text).
The bottom four panels show the binned input evolution, the
reconstructed evolution, the significance of the reconstructed
evolution and the absolute significance of the residuals.
The evidence in this case is maximised at $\rho_{\rm thresh}=2800$,
log$_{10} \, \lambda=-2.2$.}
\label{recon_real_mono}
\end{figure}

\begin{figure}
\epsfxsize=90mm
{\hfill
\epsfbox{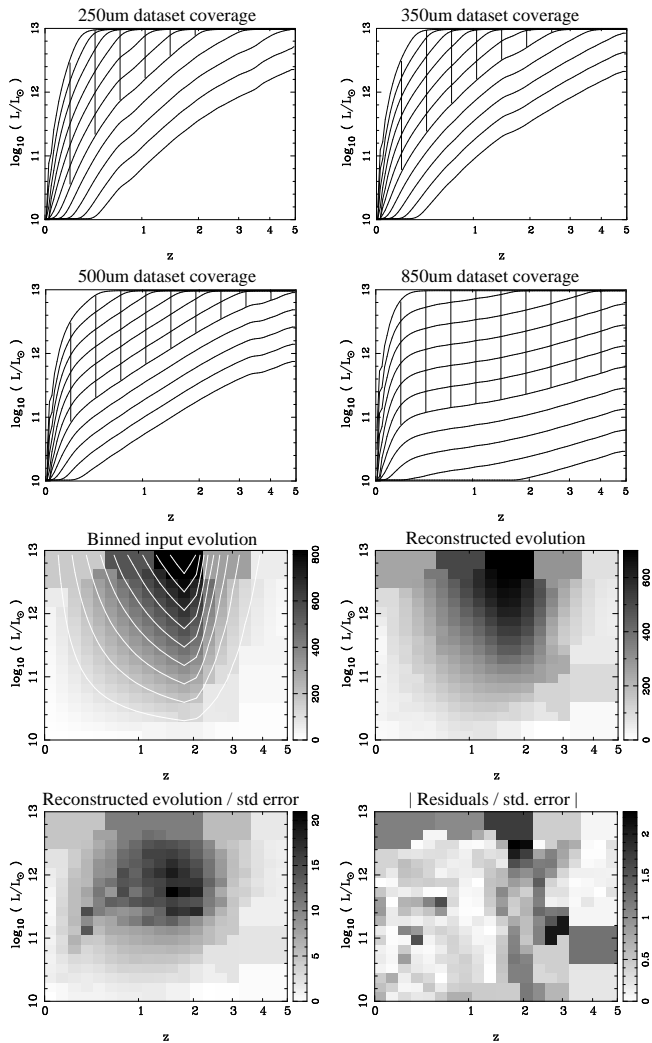}
\hfill}
\caption{Application of the reconstruction method to simulated {\sl
Herschel}/SPIRE (250, 350 and 500$\,\mu$m) and SCUBA2 (850$\,\mu$m)
data generated using the cut-off evolution function over
1$\,$deg$^2$. Panels are as described in Figure
\ref{recon_real_mono}. The evidence in this case is maximised at
$\rho_{\rm thresh}=2700$, log$_{10} \, \lambda=-2.8$.}
\label{recon_real_cutoff}
\end{figure}

We tested the method with the synthetic data generated using the two
different evolution functions as before. The top four panels of
Figures \ref{recon_real_mono} and \ref{recon_real_cutoff} show the
coverage of the datasets in the case of the monotonic and cut-off
evolution function respectively. The redshift data cover the flux
ranges 10 - 1000$\,$mJy, 20 - 1000$\,$mJy, 20 - 500$\,$mJy and 2 -
200$\,$mJy at 250, 350, 500 and 850$\,\mu$m respectively in six equal
intervals in log($L$) and 10 equal intervals in log($1+z$). Similarly,
the number count data cover four flux bins logarithmically spaced
between the ranges 0.5 - 10$\,$mJy, 1 - 20$\,$mJy, 1 - 20$\,$mJy and
0.1 - 2$\,$mJy at 250, 350, 500 and 850$\,\mu$m respectively.  We 
assumed an area of 1$\,$deg$^2$.

The bottom four panels in Figure \ref{recon_real_mono} show the
reconstruction results for the monotonically evolving data. Despite a
10 times reduction in area and lower redshift data coverage of the
$L-z$ plane compared with the high S/N case (shown in the second
column of Figure \ref{recon3}), the input evolution is recovered with
a comparable accuracy. At high luminosities, the average significance
of the residuals is very similar to that in the high S/N case. This is
not too surprising since the relative reduction in area is offset by
the more numerous redshift data across the different wavelengths.
Although the residuals fare worse at low luminosity where redshift
data are completely lacking, they are smaller than the residuals that
arise when no redshift data are present anywhere on the $L-z$ plane.
For example, the residuals at low $L$, high $z$ in the third column of
Figure \ref{recon1} are higher {\em and} evaluated over larger pixels.
The difference with the reconstruction using multi-wavelength data is
that the variation of luminosity with redshift for a given flux
changes as a function of observed wavelength. This is a direct result
of the conversion between monochromatic and bolometric luminosity
which depends on the SED assumed. The effect is that lines of constant
flux, and therefore flux bins, in the $L-z$ plane corresponding to two
different observed wavelengths can intersect each other. In the case
of the reconstruction shown in Figure \ref{recon_real_mono}, some of
the flux bins without redshift information cross other bins
corresponding to different wavelengths where redshifts are
present. Since data binned by both flux and redshift give a direct
measure of the evolution at a specific location in the $L-z$ plane,
this provides additional constraints on how the evolution must vary
along the length of a flux-only bin.

Figure \ref{recon_real_cutoff} shows how the method performs in
reconstructing the simulated {\sl Herschel}/SCUBA2 data generated
using the cut-off evolution function. As in the case of the
monotonically evolving data, the residuals exhibit a similar
level of variation. However, unlike the monotonic case, the
resolution of the recovered evolution is slightly lower than in the
high S/N case. This is especially true at the highest redshifts where
the evolution is considerably weaker than in the monotonic case. This
results in much smaller numbers of galaxies per unit area of the $L-z$
plane and hence increased Poisson noise and covariance. The adaptive
gridding procedure therefore increases the pixel size in these
regions.

\section{Summary}
\label{sec_summary}

In this paper, we have presented a new method for measuring evolution
of the galaxy luminosity function. The method computes the evolution
as a discretised function of redshift and bolometric luminosity.  The
advantage brought about by this discretisation is that the evolution
can be solved linearly, ensuring the best fit solution is always found
with a single, efficient matrix inversion. Formulating this as a
linear problem also brings the additional advantage that the
uncertainty in the reconstructed evolution can be obtained with a few
simple extra computational steps. Furthermore, by introducing linear
regularisation, the evolution can be extrapolated into regions of the
$L-z$ plane where data are lacking.

We have also developed a procedure for adaptively pixellising the
evolution function.  This allows the evolution to be reconstructed
with higher resolution in regions of the $L-z$ plane where data
constraints are stronger and lower resolution where constraints are
weaker. The procedure uses the data to automatically set the optimal
pixellisation via maximisation of the Bayesian evidence.

We have applied the method to a range of synthetic datasets
constructed with two different input evolution functions; one rising
monotonically with redshift and luminosity and the other incorporating
a redshift cutoff. Comparing the reconstructed evolution with the
input evolution provides a means of testing how well the method
performs with varying dataset type, coverage and signal-to-noise. Our
findings indicate that redshift measurements are essential to locate
the presence of a cutoff in the evolution and that these must lie
beyond the cutoff. Number count data alone allow for a surprisingly
reasonable reconstruction if the evolution function is known to be
monotonic, but falsely indicate monotonic behaviour when
a redshift cutoff is present.

We have made predictions of the degree to which forthcoming submm
surveys carried out with new instruments (SCUBA2 and Herschel)
will allow evolution in the submm GLF to be determined. These
simulations show that combining a mixture of number count data from
both facilities and including measurements of redshifts down to the
source confusion limit will allow for a very well resolved and
reliable measurement of evolution.  In particular, the predictions
have demonstrated that much improved constraints are provided by pure
number count data measured at a combination of different wavelengths,
compared to counts measured at just one wavelength.  This is a result
of the cross-linking of flux bins on the $L-z$ plane due to the
conversion between monochromatic and bolometric luminosity which is a
function of both redshift and wavelength.

In terms of applying the method to real data, there are a few
additional complexities that must be considered. One complexity is
that real data are typically incomplete. This will result in the
reconstructed evolution being a lower limit on the real evolution.
Indeed, our simulations indicate that, due to regularisation, the
reconstructed evolution tends to be underestimated even when the data
are complete, although by an amount which is within the derived
uncertainties. However, depending on the magnitude of the
incompleteness, this may give rise to an underestimate that is not
within the derived evolution error budget. In addition, depending on
the level of regularisation, severe incompleteness in one dataset
could significantly affect the reconstructed evolution in other areas
of the $L-z$ plane covered by near-complete data.

Another consideration that must be made with real data is regarding
the SED used for converting between monochromatic and bolometric
luminosity. Clearly, if this is a poor match to the SEDs of the
galaxies in a given dataset, then the assumed location of that dataset
in the $L-z$ plane will be inaccurate and give rise to a systematic
error in the recovered evolution. In the submm, this will be larger
for fluxes near the peak of the observer-frame SED, but on average,
should be a relatively small effect. Of course, instead of using a
single SED as we have done in our simulations, source-specific SEDs,
or dataset-specific SEDs could be used in practice.

In this paper, we have demonstrated the behaviour of the
reconstruction method with a combination of different datasets and
evolution functions. Although we have concentrated on key
characteristics, there are many more that could have been tested.
Since an exhaustive investigation of all configurations that might be
encountered in practice is well beyond the scope of this work, a
sensible approach would be to carry out simulations when applying the
method to datasets that differ greatly from those tested here.  In
this way, any potential systematics that might arise from the data
or any non-uniqueness in the reconstructed evolution could be
quantified.

There are also several ways in which the method could be
developed. For example, the adaptive pixellisation scheme is
relatively simple and could be enhanced to offer greater
adaptability. Another possible enhancement might be improving the use
of redshift information. In this work, we incorporated redshifts by
crudely binning number counts by redshift. This lowers the
constraining power of the redshift data. With this in mind, we have
begun development of a more sophisticated scheme that includes
redshift information on a more efficient source by source basis.  This
will be presented in forthcoming work.

We are about to enter a new era in submm cosmology with the arrival of
new instruments such as {\sl Herschel} and SCUBA2.  Surveys conducted
with these instruments will give an increase in the number of detected
submm sources of several orders of magnitude compared to existing
surveys. Their significantly improved sensitivities will allow much
wider ranges in luminosity and redshift to be explored. Combined with
new methods to measure evolution, over the next few years, this will
bring about a revolution in our overall understanding of how galaxies
form and evolve.

\begin{flushleft}
{\bf Acknowledgements}
\end{flushleft}

SD acknowledges support by the Science and Technologies Facilities
Council. We would like to thank the reviewer of this paper, Professor
Steve Phillipps, for his helpful comments and suggestions.


{}

\label{lastpage}

\end{document}